\documentstyle[preprint,aps,12pt]{revtex}

\begin{document}

\title{The General Structure of Eigenvalues of Non-linear Oscillators}
\author{A.~D.~Speliotopoulos}
\address{Higher Dimension Research, Inc., 7582 Currell Blvd. Suite 114,
St.~Paul, MN 55125} 
\date{May 24, 1998}
\maketitle

\begin{abstract} 
Hilbert Spaces of bounded one dimensional non-linear oscillators are
studied. It is shown that the eigenvalue structure of all such
oscillators have the same general form. They are dependent only on the
ground state energy of the system and a single functional $\lambda(H)$ of the
Hamiltonian $H$ whose form depends explicitly on $H$. It is
also found that the Hilbert Space of the non-linear oscillator is
unitarily {\it inequivalent\/} to the Hilbert Space of the simple
harmonic oscillator, providing an explicit example of Haag's
Theorem. A number operator for the nonlinear oscillator is constructed
and the general form of the partition function and average energy of
an non-linear oscillator in contact with a heat bath is
determined. Connection with the WKB result in the semi-classical limit 
is made. This analysis is then applied to the specific case of the 
$x^4$ anharmonic oscillator. 
\end{abstract}


\section{Introduction}

In this paper we shall study the general structure of the energy
eigenvalues for one dimensional non-linear oscillators. To be specific, we are
interested in Hamiltonians which have the form,
\begin{equation}
H = \epsilon_0 \left(a^\dagger a +\frac{1}{2}\right) + V(a,a^\dagger)\>,
\label{e1}
\end{equation}
where $V(a,a^\dagger)$ is the interaction Hamiltonian and is a
functional of $a$ and $a^\dagger$, the creation and annihilation
operators for the simple harmonic oscillator (SHO), and $\epsilon_0$
is the SHO energy scale. We shall restrict ourselves to bounding potentials
for which $V(x)\to \infty$ when $\vert x\vert\to\infty$. When $V$ is a
polynomial in $a$ and $a^\dagger$ consisting of terms
$(a^\dagger)^pa^q$ the degree $l$ of $V$ is the maximum value of
$l=p+q$ for the polynomial. As is well known, when $l\le2$ the
Hamiltonian is easily diagonalizable by either shifting the operator
by a constant (for $l=1$) or by a Bogoluibov transformation (for
$l=2$). Nonetheless, the results of these analysis have had
far reaching applications, including coherent and squeezed quantum
states $\cite{cohe}$ in quantum optics, and the theories of superfluidity
and superconductivity $\cite{fw}$.

When $l>2$ the oscillator is usually called anharmonic with the
classical example being the oscillator with an $x^4$ or
$(a+a^\dagger)^4$ interaction potential. This particular non-linear
oscillator has been extensively studied since the early 1970's
($\cite{Simon1}$- $\cite{BenWu2}$; see $\cite{Mont}$ for a review of
the literature), due mainly to the equivalence between it and the
$\phi^4$ quantum field theory in one-dimension. It is hoped 
that a detail study of this simplified system will shed some light on
the structure of $\phi^4$ theory in higher dimensions. Research on
this oscillator continues today, mainly because it provides a 
natural test bed for such approximation schemes as the
strong coupling expansion $\cite{strong}$, modified perturbations
schemes, $\cite{HS1}$, $\cite{HS2}$, variational modified 
perturbation theories $\cite{var}$, lattice methods $\cite{lat}$,
etc. More recently, Bender and Bettencourt $\cite{BenLuis}$
have provided a deeper understanding of the system by using
multiple-scale perturbation theory showing that the frequency of
oscillation depends on the energy $H$ of the state. This was
interpreted by them as an operator form of mass renormalization. 

The main purpose of this paper is not to present a new method of
calculating the energy eigenvalues of non-linear oscillators, although
we shall end up doing so. Rather, it is to study the general structure
of both the Hilbert Space and the energy eigenvalues of non-linear
oscillators with arbitrary binding $V(a,a^\dagger)$. The approach we shall
take follows most closely the analysis done for the $l\le 2$
oscillators. Namely, we shall attempt to construct, in much the same way,
operators $\tilde a$ and $\tilde a^\dagger$ from $a$
and $a^\dagger$ which diagonalizes the Hamiltonian. We find that
unlike the SHO operators, $\tilde a$ and $\tilde a^\dagger$ obey the
commutation relation $[\tilde a, \tilde a^\dagger] = \lambda(H)$ where
in general $\lambda(H)$ is a functional of $H$. Its precise form
depends on the specific choice of $H$ and is a constant only when $l
\le 2$. The study of any non-linear oscillator thereby reduces to the
study of operators having this commutation relation along with the
determination of $\lambda(H)$ and the groundstate energy of the system.   

Because $\lambda(H)$ is not a constant function in general, we find that
$\tilde a$ and $\tilde a^\dagger$ cannot be unitarily equivalent to
$a$ and $a^\dagger$. Only in the special case when $l\le 2$ does such
a transformation exists. Consequently, the Hilbert space of the
non-linear oscillator is generally {\it unitarily inequivalent\/} to
that of the simple harmonic oscillator. This is an explicit example of
Haag's Theorem, first proposed by Haag in 1955 $\cite{Haag0}$ (see also
$\cite{Haag}$) for quantum field theories. In this paper Haag actually
proved a weaker version of the theorem by showing that the unitary
transformation between the non-interacting and interacting quantum
field theories via the interaction picture does not exist. Later, this
result was extended by Hall and Wightman $\cite{Wight0}$ (see also
$\cite{Wight}$) who showed that based on the Wightman axioms the
expectation values of the product of four or fewer fields of an
interacting theory are unitarily inequivalent to those of the free theory. 

There have been other attempts at using algebraic methods to analyze
non-linear oscillators, of course, such as the action angle or time
operator methods (see for example $\cite{Gar}$-$\cite{phase}$). Both
of these methods, however, are generalization of classical analytical
techniques to quantum mechanical systems. They rely on the existence
of the phase 
$\phi$ and time $T$ operators which are canonical to the number and
Hamiltonian operators: $[N,\phi] = 1$, $[H, T] = 1$. Because of the
positivity of the spectrum of both $N$ and $H$ for bound systems, such
operators do not exist in the usual quantum mechanical system
$\cite{Pauli}$ (see, however, $\cite{Rosen}$-$\cite{Oli}$ for the
existence of such operators in extended quantum mechanical systems).
In this sense, these methods of solution are ``formal''. The approach
we have taken in this paper does not suffer from these problems. It is
not a generalization of classical techniques but is instead a
generalization of the Bogoluibov transformation and is inherently
quantum mechanical in nature. Classical solution techniques such as
the action angle are used only in the semi-classical limit where they
are expected to be valid. 

The rest of this paper is organized as follows. In {\bf Sec II} the
general Hilbert Space and energy eigenvalue structure of non-linear
oscillators are analyzed. It is found that both depend on a functional
$\lambda(H)$ of the Hamiltonian. A number operator is constructed and
the Heisenberg equations of motion are solved. Then in {\bf Sec III}
thermal or KMS states are analyzed and it is shown that both the
partition function $Z$ and and average energy $\langle H\rangle_T$ for
non-linear oscillators are similar in form to those of the SHO. In
{\bf Sec IV} a method of determining $\lambda(H)$ is outlined and in
{\bf Sec V} the connection between this method and the semi-classical
WKB result is shown. Application of this analysis to the $x^4$
potential can then be found in {\bf Sec VI}. Concluding remarks are
given in {\bf Sec VII}.

\section{General Structure}

Given a Hamiltonian $H$ constructed from $a$, and $a^\dagger$, we seek
solutions of the operator equation 
\begin{equation}
[\tilde a, H] = \epsilon_0 \lambda(H) \tilde a\>,
\label{e2}
\end{equation}
where $\tilde a$ is understood to be a functional of $a$ and $a^\dagger$.
This is an eigenvalue equation with $\tilde a$ being the ``eigenoperator''
of $H$ and $\lambda(H)$ its corresponding ``left eigenvalue'', although
unlike the standard eigenvalue equation $\lambda(H)$ is a functional of
$H$ and the ordering in eq.~$(\ref{e2})$ is
important. Eq.~$(\ref{e2})$ does not determine $\tilde a$ uniquely
since if $\tilde a$ satisfies eq.~$(\ref{e2})$, then so does
$g(H)\tilde a$ and $\tilde a g(H)$ where $g$ is any functional of
$H$. A normalization for $\tilde a$ is needed which we choose to be
\begin{equation}
H = \epsilon_0 \left({\tilde a}^\dagger \tilde a + e_g\right)\>,
\label{e3}
\end{equation}
since it diagonalizes the Hamiltonian explicitly. $\epsilon_0e_g$ is
the ground state energy of the system and is a constant. This
is very similar to the way one determines the Bogoluibov transformation
which diagonalizes the $l=2$ Hamiltonian $H=\epsilon_0 a^\dagger a +
i\epsilon_1 (a^2-(a^\dagger)^2)/2$, but now $\lambda(H)$ is a
functional of $H$. With this normalization, eq.~$(\ref{e2})$ reduces
to 
\begin{equation}
[\tilde a, \tilde a^\dagger] = \lambda(H).
\label{e4}
\end{equation}

To show that $\tilde a$ and $\tilde a^\dagger$ creates and annihilates
eigenstates of $H$, we make use of the identity,
\begin{equation}
[\tilde a, H^n] = \Big\{(\lambda(H)+H/\epsilon_0)^n -
(H/\epsilon_0)^n\Big\}\tilde a\>.
\label{e5}
\end{equation}
obtained using eq.~$(\ref{e4})$. Then for any given functional $M(H)$ which
is expandable in a Taylor series, 
\begin{equation}
[\tilde a, M(H)] = \Big\{M(\lambda(H)+H/\epsilon_0) - M(H)\Big\}\tilde a\>.
\label{e6}
\end{equation}
With this we see that if $\tilde a$ is an eigenoperator
of $H$ with left eigenvalue $\lambda(H)$, then so is $\tilde a^n$,
\begin{eqnarray}
[\tilde a^n , H] = \epsilon_0\Bigg\{ &{}&\lambda(H) +
		  \lambda(\lambda(H)+H/\epsilon_0) + 
		  \cdots +
\nonumber \\
		  &{}&\lambda(\lambda(\cdots\lambda(\lambda(H)+H/\epsilon_0)
		  \cdots+H/\epsilon_0) +H/\epsilon_0)\Bigg\} \tilde a^n\>.
\label{e7}
\end{eqnarray}
		  
Given eq.~$(\ref{e7})$, the  Hilbert Space ${\cal H}_{nl}$ for
non-linear oscillators and their energy eigenvalues are easily
constructed in much the same way as the SHO Hilbert Space ${\cal
H}_{SHO}$. Namely, if $\vert\phi\rangle_{nl}$ is an eigenstate of $H$,
then so is $\tilde a^n\vert\phi\rangle_{nl}$ as well as $(\tilde
a^\dagger)^n\vert\phi\rangle_{nl}$. Since the spectrum of the
operator $\tilde a^\dagger \tilde a$ must be  non-negative it is
straightforward to show (see {\bf Appendix A}) that if $\lambda$ is a
positive definite function then there exists a state (the groundstate)
$\vert \Omega\rangle_{nl}$ in ${\cal H}_{SHO}$ for which $\tilde
a\vert \Omega\rangle_{nl} = 0$ and has energy $\epsilon_0 e_g$
$\cite{com1}$. ${\cal H}_{nl}$ is therefore spanned by the states 
\begin{equation}
\vert n\rangle_{nl} =\frac{(\tilde a^\dagger)^n\vert\Omega
			\rangle_{nl}}{\sqrt{A_n}}\>,
\label{e8}
\end{equation}
where
\begin{eqnarray}
A_n =&{}& \lambda(e_g)\cdot
	\Bigg(\lambda(e_g)+\lambda(\lambda(e_g)+e_g)\Bigg){\cdot\>}
	\cdots{\>\cdot}  
\nonumber \\
	&{}&\Bigg(\lambda(e_g)+\lambda(\lambda(e_g)+e_g) +\cdots+
	\lambda(\lambda(\cdots\lambda(\lambda(e_g)+e_g)\cdots+
		  e_g)+e_g)\Bigg)\>.
\label{e9}
\end{eqnarray}
They are eigenstates of $H$ with eigenvalues $\epsilon_0 e_n$ where
\begin{eqnarray}
e_n &=& e_{n-1} + \lambda(e_{n-1})\>,
\nonumber \\
    &=& e_g + \lambda(e_g) + \lambda(\lambda(e_g)+e_g) + \cdots + 
		\lambda(\lambda(\cdots\lambda(\lambda(e_g)+e_g)\cdots+
		e_g)+e_g)\>.
\label{e10}
\end{eqnarray}
$\lambda$ thereby determines the splitting between successive energy
levels. 

If $\lambda(H)$ is a constant, then from eq.~$(\ref{e10})$ we see that
the energy levels of the oscillator are equally spaced. As we shall
show in the next section, this is only possible for $l\le 2$, which is
well known. When $l>2$, $\lambda(H)$ is a functional of $H$ and
this equal spacing no longer occurs.

Notice, however, that both eq.~$(\ref{e3})$ and the commutation relation
eq.~$(\ref{e4})$ are invariant under unitary unitary transformations:
$\tilde a\to U\tilde aU^\dagger$. As usual, unitary transformations
are canonical transformations which preserves the commutation
relation. For the SHO, $\lambda =1$, while for an non-linear
oscillator $\lambda(H)$ is a functional of $H$. Since a unitary
transformation cannot change the functional form of $\lambda$,  
$\tilde a$ and $a$ are unitarily inequivalent. {\it Consequently, the
Hilbert Spaces ${\cal H}_{nl}$ and ${\cal H}_{SHO}$ are unitarily
inequivalent Hilbert Spaces}. 

It is well known that under certain conditions any two solutions of
the {\it canonical commutation relation\/} $[a,a^\dagger]=1$ is connected by
an unitary transformation. Our result does not contradict this. The
operators $\tilde a$ and $\tilde a^\dagger$ which create and annihilate 
eigenstates of a general one dimensional oscillator do not in
general obey the canonical commutation relation except in the special
case of $l\le2$. Instead, $[\tilde a,\tilde a^\dagger] = \lambda(H)$.
Indeed, it is precisely for this reason that $\tilde a$ and $a$ cannot
be related to one another by an unitary transformation. 

If we consider now Hamiltonians of the form eq.~$(\ref{e1})$ in which
$V(a,a^\dagger)$ is controlled by a single coupling constant
$\epsilon_1$, we can label the Hilbert Space for each $\epsilon_1$ 
as ${\cal H}_{\epsilon_1}$. Then ${\cal H}_{\epsilon_1}$ is unitarily
inequivalent to ${\cal H}_{SHO}$. Moreover, there cannot be a unitary
transformation which maps ${\cal H}_{\epsilon_1} \to {\cal
H}_{\epsilon_1'}$ when $\epsilon_1 \ne \epsilon_1'$. If there were,
then using a succession of these transformations we can construct a
unitary transformation mapping ${\cal H}_{\epsilon_1}$ to ${\cal
H}_{SHO}$ and the two Hilbert Spaces would be unitarily
equivalent. Thus for different values of $\epsilon_1$ the Hilbert
Spaces  ${\cal H}_{\epsilon_1}$ are inequivalent to one
another. This is a concrete example of Haag's Theorem (see
$\cite{Haag}$), first proved for quantum field theories using
Lorentz invariance. 
 
Because the spectrum of $\tilde a^\dagger\tilde a$ no longer
consists of the non-negative integers, but instead depends on the
energy of the state, $\tilde a^\dagger \tilde a$ cannot in general be
interpreted as the number operator $N$ for non-linear oscillators.
We shall construct this operator explicitly by first noting
that $H$ does not change the occupation number $n$ of the states $\vert
n\rangle_{nl}$. Consequently, $[N,H]=0$ with the subsidiary 
condition that $N\vert\Omega\rangle_{nl}=0$. $N(H)$ can be a
functional of $H$ only. From eq.~$(\ref{e8})$, it is straightforward
to see that it must satisfy the commutation relation
\begin{equation}
[\tilde a,N(H)] = \tilde a\>.
\label{n1}
\end{equation}
Expanding $N(H)$ in a power series and once again using
eq.~$(\ref{e6})$, the solution to the operator equation eq.~$(\ref{n1})$
reduces to finding the solution of the {\it algebraic\/} equation
\begin{equation}
N(\lambda(e)+e)-N(e) = 1\>,
\label{n2}
\end{equation}
for a given $\lambda$ with the ``boundary condition'' $N(e_g)=0$ (see
{\bf Appendix B}). $e$ is a real number in eq.~$(\ref{n2})$ and $N(H)$
is obtained by replacing $\epsilon_0 e\to H$. (Or, equivalently,
eq.~$(\ref{n2})$ is the resultant equation after applying the
corresponding operator equation to an eigenstate of $H$ with energy
$\epsilon_0 e$.) 

Like the differential equation it resembles, the general solution of
eq.~$(\ref{n2})$ consists of the linear combination $N_p(H)+N_h(H)$
where $N_p(e)$ is the ``particular'' solution to eq.~$(\ref{n2})$
while $N_h(e)$ is the solution to the ``homogeneous'' equation
\begin{equation}
N_h(\lambda(e)+e)-N_h(e) = 0\>.
\label{n3}
\end{equation}
Whether or not such a solution exists depends on the particular form
of $\lambda(e)$, although the above equations can be solved for
general $\lambda$ in the semi-classical limit as we shall see 
in {\bf Sec.~V}. Unlike a differential equation, however, the single
boundary condition $N(e_g)=0$ is {\it not\/} sufficient to 
determine $N(H)$ uniquely in general. Consider the case of the
SHO. Then $\lambda =1$ and the particular solution of eq.~$(\ref{n2})$
gives $N_p(H) = H/\epsilon_0 - I/2$ and is, in fact, the usual number
operator. However, the solution of the homogeneous eq.~$(\ref{n3})$ 
is {\it any periodic function\/} with period $1$ which vanishes at
$e=e_g$. There are an infinite number of such functions, such as
\begin{equation}
N(H)= H/\epsilon_0 - e_gI + C \sin(\pi(H/\epsilon_0 - e_gI))\>.
\label{n4}
\end{equation}
for any real $C$.

From eqs.~$(\ref{e3})$ and $(\ref{e4})$ the study of non-linear
oscillators reduces to the determination of the groundstate energy
$e_g$ and the functional $\lambda(H)$. This is non-trivial and a
method for doing so will be given in the {\bf Sec IV}. For now we
shall limit ourselves to a qualitative description of the energy
levels by looking at different possible behaviors of $\lambda(e)$. 

For a ground state to exist,
$\lambda(e_g)>0$ and we shall restrict our considerations to such
$\lambda$. Representing the eigenvalues of $H/\epsilon_0$ generically
by $e$, if $\lambda(e)$ is a monotonically increasing function which
is unbounded from above, then the energy spacings between successive
energy levels becomes wider as $n$ increases and $e_n$ grows rapidly
with $n$. If, on the other hand, $\lambda(e)\to constant$ as $e\to
\infty$, then eventually the energy levels become equally spaced and
we would once again obtain SHO type of energy levels. Notice also that
if we consider eq.~$(\ref{e10})$ as a non-linear transformation of
$e_n$ generated by $\lambda(e)$, then the fix point of this
transformation $\lambda(\lambda(e)+e) = \lambda(e)$ occurs precisely
when $\lambda(e)$ goes to a constant (see {\bf Appendix B}). Finally,
if $\lambda(e)$ is a monotonically decreasing function of $e$ which
decreases sufficiently rapidly, there will be an upper bound to the
energy levels $e_{max}$.   

Finally, let us consider time evolution. If $H$ does not explicitly
depend on $t$, time evolution is generated by a unitary 
transformation $\cite{com2}$,
\begin{equation}
\tilde a(t) = e^{itH/\hbar}\tilde a(0)e^{-itH/\hbar}\>,
\label{e12}
\end{equation}
which preserves the commutation relation
eq.~$(\ref{e4})$. Using eq.~$(\ref{e2})$, the solution to the
Heisenberg equation of motion is
\begin{equation}
\tilde a(t) = e^{-i \epsilon_0\lambda(H)t/\hbar}\tilde a(0)\>.
\label{e15}
\end{equation}
The frequency of oscillation of $a(t)$, $\epsilon_0\lambda(H)/\hbar$,
now depends on the Hamiltonian $H$. This agrees with the recent result
of Bender and Bettencourt $\cite{BenLuis}$ and was interpreted by them as an
operator form of mass renormalization. $A_n$'s dependence in
eq.~$(\ref{e9})$ on the energy of the state would then be wavefunction
renormalization.   

\section{KMS States}

We now put the non-linear oscillator in contact with a thermal
reservoir at a temperature $T$ and consider the average energy and
number density of the system. We shall denote these thermal averages
by $\langle \cdots\rangle_T$ which we shall take to be a KMS state
$\cite{K}$-$\cite{Haag2}$. Namely, if $A(t)$ and $B(t)$ are two
operators in the Heisenberg representation, then
\begin{equation}
\langle A(t)B(t)\rangle_T = \langle B(t)A(t+i\hbar\beta)\rangle_T\>,
\label{t1}
\end{equation}
where $1/\beta = k_B T$. Since thermal equilibrium states are
stationary, we can without a loss of generality take $t=0$ in
eq.~$(\ref{t1})$. 

Applying this condition to eq.~$(\ref{e4})$,
\begin{equation}
\langle\lambda(H)\rangle_T = \langle\tilde a(0)\tilde a^\dagger(0)\rangle_T
				-
			     \langle\tilde a(0)\tilde
				a^\dagger(i\hbar\beta)\rangle_T\>.
\label{t2}
\end{equation}
Then using the solution eq.~$(\ref{e15})$ of the Heisenberg equation
of motion, eq.~$(\ref{e2})$ and the commutation relation eq.~$(\ref{e4})$, 
\begin{equation}
\langle\lambda(H)\rangle_T = \left\langle\left(H/\epsilon_0 -e_g\right)
			\left(1-e^{-\epsilon_0\beta\lambda(H)}\right)
			\right\rangle_T\>,
\label{t3}
\end{equation}
which reduces to the usual Bose-Einstein distribution for the SHO when
$\lambda=1$. 

Unlike the case of the SHO it is not possible to determine $\langle
H\rangle_T$ any further using solely the KMS condition. We must
make use of a partition function and shall restrict ourselves to
states which can be represented by a trace over a density matrix,
\begin{equation}
\langle H \rangle_T = \frac{1}{Z}{\rm\bf Tr}_{{\cal H}_{nl}} \>H
			e^{-\beta H}\>,
\label{t4}
\end{equation}
where $Z\equiv {\rm\bf\> Tr}_{{\cal H}_{nl}} \>e^{-\beta H}$ is the usual
partition function. Then using the identity
\begin{equation}
\langle
[H+\epsilon_0\lambda(H)]e^{-\epsilon_0\beta\lambda(H)}\rangle_T =
	\langle e^{-\epsilon_0\beta\lambda(H)}\rangle_T \langle H\rangle_T 
	-
	\frac{\partial\>\>\>}{\partial\beta}\langle
	e^{-\epsilon_0\beta\lambda(H)}\rangle_T\>, 
\label{t5}
\end{equation}
we obtain
\begin{equation}
\langle H\rangle_T = \epsilon_0 e_g - 
		     \frac{
			\frac{\partial\>\>\>}{\partial\beta}
			\langle e^{-\epsilon_0\beta\lambda(H)}\rangle_T}
		     {1-\langle e^{-\epsilon_0\beta\lambda(H)}\rangle_T}\>,
\label{t6}
\end{equation}
and we see once again the importance of $\lambda(H)$. Indeed, from
eq.~$(\ref{t4})$ we find that
\begin{equation}
Z = \frac{e^{-\beta\epsilon_0
e_g}}{1-\langle e^{-\epsilon_0\beta\lambda(H)}\rangle_T}\>.
\label{t7}
\end{equation}

As for the number operator, from eq.~$(\ref{n2})$,
\begin{equation}
\langle e^{-\epsilon_0\beta\lambda(H)}\rangle_T = 
	\langle N(H+\epsilon_0\lambda(H))e^{-\epsilon_0\beta\lambda(H)}\rangle_T
	-
	\langle N(H)e^{-\epsilon_0\beta\lambda(H)}\rangle_T\>.
\label{t8}
\end{equation}
Then using
\begin{equation}
\langle N(H+\epsilon_0\lambda(H))e^{-\epsilon_0\beta\lambda(H)}\rangle_T
	=\frac{1}{Z}{\rm\bf Tr}_{{\cal H}_{nl}}\>
	N(H+\epsilon_0\lambda(H))e^{-\beta H-\epsilon_0\beta\lambda(H)}\>, 
\label{t9}
\end{equation}
and eq.~$(\ref{e10})$, we find that $\langle
N(H+\epsilon_0\lambda(H))e^{-\epsilon_0\beta\lambda(H)}\rangle_T =
\langle N(H)\rangle_T$ so that
\begin{equation}
\langle e^{-\epsilon_0\beta\lambda(H)}\rangle_T = \left\langle
N(H)\left(1-e^{-\epsilon_0\beta\lambda(H)}\right)\right\rangle_T\>.
\label{t10}
\end{equation}
This once again agrees with the SHO result for $\lambda=1$.

\section{Solution of the Eigenvalue Problem}

$\tilde a$ and $\lambda(H)$ can be determined in the following
manner. Since $H$ is given in terms of $a$ and $a^\dagger$, in general
$\tilde a = \tilde a(a, a^\dagger)$, which is understood in terms of a
power series,
\begin{equation}
\tilde a = \sum_{r,s=0}^\infty b_{rs} (a^\dagger )^r a^s\>.
\label{e16}
\end{equation}
By using the commutation relation $[a,a^\dagger]=1$, we can
always reduce any expansion of $\tilde a$ to this
form. Eq.~$(\ref{e16})$ is well defined only if the corresponding
function  
\begin{equation}
f(z, \bar z) = \sum_{r,s=0}^\infty b_{rs} \bar z^r z^s\>,
\label{e17}
\end{equation}
is absolutely convergent on ${\bf R}^2$. 

At this point we should also express $\lambda(H)$ as a power
series in $H/\epsilon_0$, insert this series and as well as
eq.~$(\ref{e16})$ in eq.~$(\ref{e2})$ and obtain an infinite set of
coupled equations between various $b_{rs}$ and the coefficients of the
$\lambda$ expansion. The problem would quickly become intractable,
however. We shall therefore first make the following drastic 
simplification. Instead of eq.~$(\ref{e2})$ we shall solve the simpler
equation 
\begin{equation}
:[\tilde a, H]: = \epsilon_0\lambda(H)\tilde a\>,
\label{e18}
\end{equation}
where $:\>:$ denotes normal ordering. Correspondingly, we shall take the
normalization condition as
\begin{equation}
:H: = \epsilon_0\left(:\tilde a^\dagger\tilde a:+\frac{1}{2}\right)\>.
\label{e19}
\end{equation}
We shall then use the solution of this equation as a guide to
reconstructing the solution to eq.~$(\ref{e2})$. Notice that
corrections to the groundstate energy cannot be determined under this
simplification and can only be determined when the full operators are
reconstructed from the solution to eq.~$(\ref{e18})$. 
 
Denoting the solution to eq.~$(\ref{e18})$ by the superscript $sc$, we
find that for
\begin{equation}
\tilde a^{(sc)} = \sum_{r,s=0}^\infty b_{rs}^{(sc)} (a^\dagger )^r a^s\>,
\label{e20}
\end{equation}
we have 
\begin{eqnarray}
\epsilon_0\sum_{r,s=0}^\infty b_{rs}^{(sc)}\lambda(H)(a^\dagger )^r a^s 
	= \sum_{r,s=0}^\infty  
		b_{rs}^{(sc)}\Bigg\{&{}&\epsilon_0 (s-r)(a^\dagger )^r a^s
		+
	s:(a^\dagger)^r a^{s-1}[a, V(a,a^\dagger)]: + 
\nonumber \\
	 &{}&
	r:(a^\dagger)^{r-1} [a^\dagger, V(a,a^\dagger)]a^s:\Bigg\}
	\>.
\label{e21}
\end{eqnarray}
Under this normal ordering, solving eq.~$(\ref{e21})$ is
equivalent to solving the differential equation,
\begin{equation}
\lambda^{(sc)}\left(e^{(sc)}\right) f^{(sc)}(z,\bar z)
			      =\{e^{(sc)},f^{(sc)}\}_{PB}\equiv 
			      \frac{\partial
			      e^{(sc)}}{\partial\bar z} 
			      \frac{\partial f^{(sc)}}{\partial z}
				-
			      \frac{\partial e^{(sc)}}{\partial z}
			      \frac{\partial f^{(sc)}}{\partial
			      \bar z}\>,
\label{e22}
\end{equation}
where $\tilde a^{(sc)} = f^{(sc)}(a,a^\dagger)$. $e^{(sc)}$ is obtained from
$H/\epsilon_0$ by replacing everywhere $a \to z$ and $a^\dagger \to
\bar z$ and the normalization condition $(\ref{e19})$ is now $e^{(sc)} = \vert
f^{(sc)}\vert^2 + 1/2$. The right hand side of eq.~$(\ref{e22})$ is just the
classical Poisson bracket but with the generalized coordinates  
\begin{equation}
z = \left(\frac{m\epsilon_0}{2\hbar^2}\right)^{1/2}x +
i\left(\frac{1}{2m\epsilon_0}\right)^{1/2}p 
\qquad,\qquad  
\bar z = \left(\frac{m\epsilon_0}{2\hbar^2}\right)^{1/2}x -
i\left(\frac{1}{2m\epsilon_0}\right)^{1/2}p\>. 
\label{e23}
\end{equation}
where $m$ is the mass of the particle. 
We are therefore looking for a semi-classical solution to
eq.~$(\ref{e3})$. Indeed, we shall see explicitly in {\bf Sec. V} that
the solution of eq.~$(\ref{e22})$ is equivalent to the WKB approximation.

Importantly, eq.~$(\ref{e22})$ has the same symmetry properties as
eq.~$(\ref{e2})$. Namely, if $f^{(sc)}(z,\bar z)$ is a solution to
eq.~$(\ref{e22})$, then so is $f^{(sc)}(z,\bar z) g(e^{(sc)})$ where 
$g(e^{(sc)})$ is any function of $e^{(sc)}$ (although they do not
satisfy the normalization condition eq.~$(\ref{e19})$). Making use of this
symmetry, we change coordinates to $e^{(sc)}$ and $\theta
=-i\log(z/\bar z)/2$ from $\vert z\vert$ and $\theta$. Then
eq.~$(\ref{e22})$ reduces to
\begin{equation}
i\lambda^{(sc)}\left(e^{(sc)}\right)
	f^{(sc)}\left(e^{(sc)},\theta\right) = \frac{\partial 
	e^{(sc)}}{\partial \vert z\vert^2} \frac{\partial f^{(sc)}}{\partial
	\theta}\>, 
\label{e25}
\end{equation}
whose solution is
\begin{equation}
f^{(sc)}(e^{(sc)},\theta) = \sqrt{e^{(sc)}-1/2}
	\exp\left\{i\lambda^{(sc)}\int^\theta_0  
	\left(\frac{\partial e^{(sc)}}{\partial \vert
	z\vert^2}\right)^{-1}d\phi\right\}
\label{e26}
\end{equation}
and satisfies the normalization condition explicitly. Determination
of $\tilde a^{(sc)}$ is then reduced to performing the above
integral, which requires inverting the equation $e^{(sc)} =
e^{(sc)}(\vert z\vert, \theta)$ and solving for $\vert z\vert$ in terms of
$e^{(sc)}$ and $\theta$. Next, for $f^{(sc)}$ to be analytic on
${\bf R}^2$, $f^{(sc)}\left(e^{(sc)},0\right) =
f^{(sc)}\left(e^{(sc)},2\pi\right)$, giving 
\begin{equation}
\frac{1}{\lambda^{(sc)}\left(e^{(sc)}\right)} = \frac{1}{2\pi}\int^{2\pi}_0 
	\left(\frac{\partial e^{(sc)}}{\partial \vert
	z\vert^2}\right)^{-1}d\phi\>,
\label{e27}
\end{equation}
which determines $\lambda^{(sc)}$.
Notice that in contrast to phase angle techniques which require the
construction of a phase {\it operator\/} (see, for example
$\cite{phase}$) and its concombinant difficulties, our analysis uses
the phase only in the semi-classical limit where it is well defined.
 
Reconstruction of $\tilde a$ and $\lambda(H)$ from $f^{(sc)}(z,\bar
z)$ is now straightforward, although tedious. $\tilde a^{(sc)}$ can be
obtained by first expanding $f^{(sc)}$ in eq.~$(\ref{e26})$ in a power
series in $z$ and $\bar z$, then taking $\tilde a^{(sc)} =
f^{(sc)}(a,a^\dagger)$. Since $\tilde a^{(sc)}$ was obtained via
normal ordering, there is an ordering ambiguity when we reconstruct
$\tilde a$ from it. Fundamentally, this arises when we replace
$z\to a$, $\bar z\to a^\dagger$ in $f^{(sc)}(a,a^\dagger)$ since the
term $\bar z z$ in the expansion can be replaced by either $a^\dagger
a$ or $a a^\dagger = 1+a^\dagger a$. Therefore, to determine $\tilde
a$ we shall take $\tilde a =
\left\{f^{(sc)}(a,a^\dagger)\right\}_{order}$, but 
we now replace $\bar z z \to a^\dagger a + A$ where the $A$'s are
constants. These are determined by requiring that the resulting
expansions for $\tilde a$ and $\lambda(H)$ satisfy both
eq.~$(\ref{e2})$ and eq.~$(\ref{e3})$ (or equivalently
eq.~$(\ref{e4})$) term by term in the expansion. This uniquely
determines not only $\tilde a$ and $\lambda(H)$, but $e_g$ as well. 

From eq.~$(\ref{e27})$ we see that for $\lambda(H)$ to be independent
of $H$, $\partial e^{(sc)}/\partial \vert z\vert^2 = k$, where $k$ is a
function of $\theta$ only. This limits $l\le 2$. Correspondingly, if
$l>2$, $\lambda(H)$ is necessarily a functional of $H$. 

\section{The WKB Approximation}

We now make the connection between the solution of eq.~$(\ref{e22})$
and the semi-classical limit. From the correspondence principle, in
the large $n$ limit $e_n$ goes over to the classical result. The spacings
between energy levels $e_n-e_{n-1}$ are small in comparison to
$e_{n-1}$ and the levels are essentially continuous. In this limit, we
can then approximate 
\begin{equation}
e_n - e_{n-1} \approx \frac{de}{dn}\>,
\label{w1}
\end{equation}
where $e(n)$ considered as a continuous function of $n$. Then from
eq.~$(\ref{e10})$ 
\begin{equation}
\frac{de}{dn} \approx \lambda^{(sc)}(e)\>,
\label{w2}
\end{equation}
where we have replaced $\lambda\to\lambda^{(sc)}$ in this
limit. Integrating and using eqs.~$(\ref{e23})$ and $(\ref{e27})$,
\begin{equation}
n +n_\infty \approx \frac{1}{2\pi}\int_0^e\int_0^{2\pi} 
		    \left(\frac{\partial
		    e^{(sc)}}{\partial\vert z\vert^2}\right)^{-1}de^{(sc)}d\theta\>,
\label{w3}
\end{equation}
where $n_\infty$ is an integration constant which can be neglected in the
limit $n\to\infty$. Changing variables back to $z$ and $\bar z$ in
the integrand of eq.~$(\ref{w3})$ and using eq.~$(\ref{e23})$, we find
that 
\begin{equation}
n +n_\infty \approx \frac{1}{2\pi\hbar}\int\int_{{\cal D}_{\epsilon_0
e}}dx\,dp\>. 
\label{w4}
\end{equation}
The integration is now over a disk ${\cal D}_{\epsilon_0 e}$ centered
about the origin in the classical phase space. This is just the semi-classical
Bohr-Sommerfeld quantization rule obtained from the WKB approximation. 

We next consider the solution of the algebraic eq.~$(\ref{n2})$ in
the large $n$ limit. Since $\lambda(e)$ measures the energy splitting
between energy levels, in this limit $\lambda(e)\ll e$ (see {\bf
Appendix A}) and eq.~$(\ref{e9})$ can be approximated by the
differential equation 
\begin{equation}
\lambda^{(sc)}(e)\frac{dN}{de} \approx 1\>,
\label{w5}
\end{equation}  
where once again we have replaced $\lambda\to\lambda^{(sc)}$. The
solution to this equation is trivial and we once again obtain the WKB
result, 
\begin{equation}
N(e) +n_\infty \approx \frac{1}{2\pi\hbar}\int\int_{{\cal D}_{\epsilon_0
e}} dxdp\>.
\label{w6}
\end{equation}
Notice, however, that now the {\it operator} $N(H)$ can now be
obtained directly from eq.~$(\ref{w6})$ by expanding the integral in
powers of $e$ and replacing $e\to H/\epsilon_0$.

Finally, we consider the quantum partition function 
\begin{equation}
Z \equiv {\rm Tr}_{{\cal H}_{nl}} e^{-\beta H}= \sum_{n=0}^{\infty}
e^{-\beta e_n}\>, 
\label{w7}
\end{equation}
in the large temperature limit. Making use of the Euler-Maclaurin
formula, 
\begin{equation}
Z \approx \int_{0}^\infty e^{-\beta\epsilon_0 e} dn + 
		\frac{1}{2}e^{-\beta\epsilon_0} + {\cal
		O}\left(e^{-\beta\epsilon_0}\right)\>.
\label{w8}
\end{equation}
In the large temperature limit $k_BT\gg \epsilon_0e_g$ we can neglect
the terms $\sim e^{-\beta\epsilon_0}$. Moreover, at this energy scale
$k_BT$, $e_n\gg \lambda(e_n)$. Then using eq.~$(\ref{w2})$, we convert
the integral over $n$ to one over $e$. Making use once again of
eq.~$(\ref{e23})$, we find that
\begin{equation}
Z \approx \frac{1}{2\pi\hbar}\int\int e^{-\beta\epsilon_0 e} dx\,dp\>,
\end{equation}
where the integral is over the classical phase space. This is
precisely the classical result with the requisite factor of 
the fundamental phase space volume $2\pi \hbar$.

\section{The $x^4$ Interaction} 

In this section we shall apply the above analysis to a non-trivial
system: the $x^4$ anharmonic oscillator,
\begin{equation}
H = \epsilon_0 \left(a^\dagger a +\frac{1}{2}\right)+
\frac{\epsilon_1}{4} (a + a^\dagger)^4\>, 
\label{e28}
\end{equation}
which corresponds to 
\begin{equation}
e^{(sc)} = \frac{1}{2}+\vert z\vert^2 + \frac{4\epsilon_1}{\epsilon_0} \vert
z\vert^4 \cos^4\theta\>.
\label{e29}
\end{equation}
Then
\begin{equation}
f^{(sc)}\left(e^{(sc)}, \theta\right) = \sqrt{e^{(sc)}-1/2} \exp\left\{
	\frac{\pi i}{2}
	\frac{I(\theta)}{I(\pi/2)}
	\right\}\>, 
\label{e30}
\end{equation}
where
\begin{equation}
I(\theta) = \int^\theta_0 \frac{d\phi}{\sqrt{1+\xi\cos^4\phi}}\>,
\label{e31}
\end{equation}
and $\xi = 16 (e^{(sc)}-1/2)\epsilon_1/\epsilon_0$. When $\epsilon_1>0$, this
integral can be reduced to 
\begin{equation}
I(\theta) = \frac{1}{2(1+\xi)^{1/4}}F(\alpha\vert q)
\label{e32}
\end{equation}
where $F(\alpha\vert q)$ is the elliptical integral of the first kind and
\begin{equation}
\alpha = {\rm arccos}\left(\frac{\sqrt{1+\xi}-\tan^2\theta}
	{\sqrt{1+\xi}+\tan^2\theta}\right)\>,
\label{e33}
\end{equation}
while
\begin{equation}
q = \frac{\sqrt{1+\xi}-1}{2\sqrt{1+\xi}}\>,
\label{e34}
\end{equation}
is its modulus. The analyticity of
$f^{(sc)}\left(e^{(sc)},\theta\right)$ gives 
\begin{equation}
\lambda^{(sc)}\left(e^{(sc)}\right) =
\frac{\pi}{2}\frac{(1+\xi)^4}{K(\sqrt q)}\>, 
\label{e35}
\end{equation}
where $K(\sqrt q)$ is the quarter period of $F(\alpha\vert q)$.

Although our analysis is strictly valid only when $\epsilon_1>0$, it
is instructive to see what happens for $\epsilon_1<0$. For
$\vert\xi\vert \le1$,  
\begin{equation}
I(\theta) = \frac{1}{\sqrt{1+\sqrt{\vert\xi\vert}}}F(\alpha'\vert q')\>,
\label{e36}
\end{equation}
where now
\begin{equation}
\alpha' =  {\rm arctan}\left(\frac{\tan\theta}
	  		{\sqrt{1+\sqrt{\vert\xi\vert}}}\right)\>,
\label{e37}
\end{equation}
and
\begin{equation}
q' = \frac{
	2\sqrt{
		\vert\xi\vert
	      }
	}
	{
	1+\sqrt{\vert\xi\vert}}\>.
\label{e38}
\end{equation}
Then
\begin{equation}
\lambda^{(sc)}(e) = \frac{\pi}{2}
		    \frac{\sqrt{1+\sqrt{\vert\xi\vert}}}{K(\sqrt{q'})}\>,
\label{e39}
\end{equation}
which vanishes when $\vert\xi\vert = 1$.

When $\vert\xi\vert>1$, $I(\theta)$ is complex and
$f^{(sc)}$ no longer satisfies the normalization condition
eq.~$(\ref{e19})$. $\lambda^{(sc)}$ is ill-defined. Consequently, the
energy states are bounded, as is well known, by $e <1/2+
\epsilon_0/(16\vert\epsilon_1\vert)$.

To determine $\tilde a$, we expand eq.~$(\ref{e30})$ to third order in
$\vert z\vert^3$, 
\begin{eqnarray}
f^{(sc)}(z,\bar z) = &z& + \frac{1}{4}\frac{\epsilon_1}{\epsilon_0}
		   \left\{-3(z^2-\bar z^2)z + (z+\bar z)^3\right\}
\nonumber \\
		  &+&
		  \frac{1}{2}\left(\frac{\epsilon_1}{\epsilon_0}\right)^2
		  \left\{
			\frac{3}{2}z^5 +\frac{39}{4}\bar z z^4 
			-\frac{25}{8}\bar z^2 z^3 -12\bar z^3 z^2
			-\frac{3}{8}\bar z^4 z +\frac{1}{4}\bar z^5
		  \right\}\>.
\label{e40}
\end{eqnarray}
We then replace $z\to a$ and $\bar z\to
a^\dagger$ in the above and take
\begin{equation}
\tilde a = a + \frac{1}{4}\frac{\epsilon_1}{\epsilon_0}F
		  +\frac{1}{2}\left(\frac{\epsilon_1}{\epsilon_0}\right)^2 G\>,
\label{e41}
\end{equation}
where
\begin{eqnarray}
F = &{}&-3(a^2-(a^\dagger)^2)a + (a+a^\dagger)^3 + f_1 a + f_2
	a^\dagger\>,
\nonumber \\
G = &{}&\frac{3}{2}a^5 +\frac{39}{4}a^\dagger a^4
		-\frac{25}{8}(a^\dagger)^2 a^3 
		-12(a^\dagger)^3 a^2 -\frac{3}{8}(a^\dagger)^4 a
		+\frac{1}{4}(a^\dagger)^5 
\nonumber \\
    &{}&	+ g_1 a^3 + g_2 a^\dagger a^2 + g_3(a^\dagger)^2 a +
		g_4 (a^\dagger)^3 + g_5 a + g_6 a^\dagger \>.
\label{e42}
\end{eqnarray}
The constants $f_1$, $f_2$, $g_1-g_6$ are present due to the
ordering ambiguity. Requiring that eq.~$(\ref{e41})$ satisfies
eq.~$(\ref{e2})$ gives $f_1 = f_2 =3$, while 
\begin{equation}
g_1+g_3 = -15\quad,\quad g_2 = -\frac{135}{8}\quad,\quad g_4 =
-\frac{3}{8}\quad,\quad g_5 = \frac{-153}{8}\quad,\quad g_6 = -
\frac{27}{2}\>.
\label{e43}
\end{equation}
The groundstate energy of the oscillator is also determined to this order,
\begin{equation}
e_g = \frac{1}{2}+\frac{3}{4}\frac{\epsilon_1}{\epsilon_0} -
\frac{21}{8}\left(\frac{\epsilon_1}{\epsilon_0}\right)^2 \>.
\label{e44}
\end{equation}
Using now the commutation relation eq.~$(\ref{e4})$, we obtain
\begin{equation}
\lambda(H) = I 	     +
	     3\left(\frac{\epsilon_1}{\epsilon_0}\right)
	     \left(\frac{H}{\epsilon_0} + \frac{I}{2}\right)
	     - \left(\frac{\epsilon_1}{\epsilon_0}\right)^2 
	     \left\{\frac{69}{4}
		\left(\frac{H}{\epsilon_0}+\frac{I}{2}\right)^2 -
		\frac{9}{2}\left(\frac{H}{\epsilon_0}+\frac{I}{2}\right)
		+ \frac{15}{2}\right\}\>,
\label{e45}
\end{equation}
while
\begin{equation}
3g_1 + g_3 = \frac{45}{2}\>,
\label{e46}
\end{equation}
giving $g_1 = 75/4$ and $g_3 = -135/4$. This last relationship was
obtained by requiring that $\lambda$ is a function of $H$ only. 
To this order then,
\begin{eqnarray}
\tilde a = a &+& \frac{1}{4}\frac{\epsilon_1}{\epsilon_0}\Bigg\{
		-3(a^2-(a^\dagger)^2)a + (a+a^\dagger)^3 + 3(a+a^\dagger)
		\Bigg\}+
\nonumber \\
	     	&{}&\frac{1}{2}\left(\frac{\epsilon_1}{\epsilon_0}\right)^2
		\Bigg\{
		\frac{3}{2}a^5 +\frac{39}{4}a^\dagger a^4
		-\frac{25}{8}(a^\dagger)^2 a^3 
		-12(a^\dagger)^3 a^2 -\frac{3}{8}(a^\dagger)^4 a
		+\frac{1}{4}(a^\dagger)^5 +
\nonumber\\    		
		&{}&
		\frac{75}{4} a^3 - \frac{135}{8} a^\dagger a^2 -
		\frac{135}{4}(a^\dagger)^2 a -
		\frac{3}{8} (a^\dagger)^3 -\frac{153}{8} a -
		\frac{27}{2} a^\dagger \>
		\Bigg\}\>.
\label{e47}
\end{eqnarray}

The energy levels can now be straightforwardly calculated from
eq.~$(\ref{e10})$, 
\begin{eqnarray}
e_n &=& e_{n-1} + \lambda(e_{n-1})\>,
\nonumber \\
    &=& e_g + \sum^{n-1}_{r=0} \lambda(e_r)\>.
\label{e48}
\end{eqnarray}
Using eq.~$(\ref{e45})$, and keeping terms to order
$(\epsilon_1/\epsilon_0)^2$ only, we obtain after re-arrangement,
\begin{eqnarray}
e_n = &e_g& + n + 3\frac{\epsilon_1}{\epsilon_0}\sum^{n-1}_{r=0}(r+1)
\nonumber \\
	&-&\left(\frac{\epsilon_1}{\epsilon_0}\right)^2
	\left(
	\frac{51}{4}\sum^{n-1}_{r=0}r^2 +
	\frac{51}{2}\sum^{n-1}_{r=0}r + 18n+\frac{21}{8}
	\right)\>. 
\label{e49}
\end{eqnarray}
Then
\begin{equation}
e_n =  n + \frac{1}{2} +\frac{3}{4}\frac{\epsilon_1}{\epsilon_0}(2n^2+2n+1)
	-\left(\frac{\epsilon_1}{\epsilon_0}\right)^2
	\left(
	\frac{17}{4}n^3+\frac{51}{8}n^2 + \frac{59}{8}n+\frac{21}{8}
	\right)\>,
\label{e50}
\end{equation}
which is the standard second order perturbation result. Notice also
that if we keep terms only up to $\epsilon_1/\epsilon_0$, then $e_n-e_{n-1}
\equiv \lambda(e_n)\approx 1 + 3n\epsilon_1/\epsilon_0$. This is
precisely the result obtained by Bender and Bettencourt
$\cite{BenLuis}$.

It is instructive to compare eq.~$(\ref{e45})$ with the expansion of
eq.~$(\ref{e35})$, 
\begin{equation}
\lambda^{(sc)}(e) = 1 + 3
	\left(\frac{\epsilon_1}{\epsilon_0}\right)(e^{(sc)}-1/2) 
	-\frac{69}{4}\left(\frac{\epsilon_1}{\epsilon_0}\right)^2(e^{(sc)}-1/2)^2\>.   
\label{e51}
\end{equation}
Notice that in both expansions the coefficients of the highest power
of the energy in each term are the same. This is a generic
feature. Quantum mechanical corrections to $\lambda^{(sc)}$ only
results in the appearance of lower powers of $H/\epsilon_0$ in each
term of the expansion. Moreover, if we then use $\lambda^{(sc)}$ to
calculate $e_n$, we find 
\begin{equation}
e_n^{(sc)} =  n + \frac{1}{2} +\frac{3}{2}\frac{\epsilon_1}{\epsilon_0}(n^2-n)
	-\left(\frac{\epsilon_1}{\epsilon_0}\right)^2
	\left(
	\frac{17}{4}n^3-\frac{33}{8}n^2 - \frac{1}{8}n
	\right)\>,
\label{e52}
\end{equation}
which also agrees with eq.~$(\ref{e48})$ in the large $n$ limit. This
also is a generic feature of the expansion since the coefficient of the
highest power of $n$ in each term of the expansion is obtained from
$\lambda^{(sc)}$ only.

The above perturbative result is valid only for small $\epsilon_1$ and
$n$. In the large $n$ limit, the semi-classical result is valid
and
\begin{equation}
\lambda(e_n)\approx \lambda^{(sc)}(e_n)
	   \approx \frac{\pi}{K(1/\sqrt 2)}\left(\frac{e_n\epsilon_1}
	   {\epsilon_0}\right)^{1/4}\>.
\label{e53}
\end{equation} 
To compare with the WKB result, from $\cite{Mont}$ we know that 
\begin{equation}
e_n^{WKB} \approx \frac{3^{4/3}\pi^2}{\left[\Gamma(1/4)\right]^{8/3}}
		\left(\frac{\epsilon_1}{\epsilon_0}\right)^{1/3}n^{4/3}\>.
\label{e54}
\end{equation}
This gives the energy splitting between levels as
\begin{equation}
e_{n+1}^{WKB} - e_n^{WKB} \approx \frac{4}{3}\frac{e^{WKB}}{n}
	\approx \frac{4\pi^{3/2}}{\left[\Gamma(1/4)\right]^2}
		 \left(\frac{e_n^{WKB}\epsilon_1}{\epsilon_0}\right)^{1/4}\>.
\label{e55}
\end{equation}
Since $K(1/\sqrt 2) = \left[\Gamma(1/4)\right]^2/4/\sqrt{\pi}$, this is
precisely the form of $\lambda(e_n)$ for large $n$ and we see
explicitly the equivalence between $\lambda^{(sc)}$ and the WKB
approximation.  

\section{Discussion}

We have shown that the study of non-linear oscillators is equivalent
to the study of algebras satisfying eqs.~$(\ref{e3})$ and
$(\ref{e4})$; the SHO being a special case of this algebra. In addition,
the Hilbert Space eq.~$(\ref{e8})$ and eigenvalues eq.~$(\ref{e10})$ 
of these algebras all have the same form. The number operator for
non-linear oscillators was also constructed. Results of this general 
analysis were used to determine the general form of the partition
function and average energy for an non-linear oscillator in contact
with a heat bath. Analysis of non-linear oscillators therefore reduces to
determining the function $\lambda(H)$ and the groundstate $e_g$ of the
oscillator. This can be done by first making a semi-classical
approximation, which requires only the evaluation of a single
integral, and then using it as a guide to constructing $\tilde a$ and
$\tilde a^\dagger$ in terms of $a$ and $a^\dagger$. This analysis was
applied to the $x^4$ interaction and both the standard second order
perturbation result as will as the WKB result were obtained. Moreover,
the recent results of Bender and Bettencourt were also obtained within
this framework.   

Unlike the Bogoluibov transformation, the mapping between $(\tilde a,
\tilde a^\dagger)$ and $(a,a^\dagger)$ is non-linear and cannot be
generated by a simple unitary transformation. The two Hilbert Spaces
${\cal H}_{\epsilon_1}$ and ${\cal H}_{SHO}$ are unitarily
inequivalent. Indeed, each value of $\epsilon_1$ determine
separate Hilbert Spaces all of whom are inequivalent to one
another. This result provides a concrete example of Haag's Theorem
proved first for quantum field theories in higher dimensions. Based on
the results of this theorem and the generality of our analysis, we
expect a similar construction to hold for the $\phi^4$ theory in
higher dimensions. Notice, however, that this construction requires a
natural energy scale to define $\lambda(H)$. For Hamiltonians of the
form eq.~$(\ref{e1})$ we have such an energy scale: $\epsilon_0$. For
quantum field theories, however, no such natural energy scale
exists. An energy scale would have to be introduced, providing a
natural introduction of a high (or low) energy cutoff in the theory.

\centerline{\bf Appendix A}

The proof that a ground state exists for the non-linear oscillator
when $\lambda(e)>0$ for $e\ge e_g$ follows in the same manner as that
for the SHO. Let $\hat H = H/\epsilon_0 - e_g = {\tilde a}^\dagger
\tilde a$. Then, if $\vert \phi_0\rangle$ is an eigenstate of $\hat H$ with
eigenvalue $\phi_0$,
\begin{equation}
\phi_0 = \langle \phi_0\vert \hat H\vert \phi_0\rangle 
       = \Big\vert \tilde a\vert\phi_0\rangle\Big\vert^2\ge 0\>,
\label{a1}
\end{equation}
and all the eigenvalues of $\hat H$ are non-negative. Next, consider the
state $\vert\phi_{-1}\rangle \equiv \tilde a \vert \phi_0\rangle$. Then
\begin{equation}
\hat H \vert \phi_{-1}\rangle = \phi_0 \vert\phi_{-1}\rangle -
\lambda(H)\vert\phi_{-1}\rangle\>.
\label{a2}
\end{equation}
Since $\lambda$ is a functional of $\hat H + e_g$, from the above
$\vert\phi_{-1}\rangle$ must be an eigenvalue of $\hat H$ also,
which we shall label as $\phi_{-1}$. Then eq.~$(\ref{a2})$ reduces to 
$\phi_{-1} = \phi_0 -\lambda(\phi_{-1}+e_g)$. Since $\lambda(e)>0$, we
have $\phi_{-1}<\phi_0$. 

Similarly, the states $\vert\phi_{-n}\equiv \tilde a^n\vert\phi_0\rangle$
are also eigenstates of $\hat H$ with eigenvalues
$\phi_{-n}$. Moreover, they satisfy a sequence of strict inequalities 
\begin{equation}
\phi_{-n} < \phi_{-(n-1)} < \cdots <\phi_{-1} < \phi_0\>.
\label{a3}
\end{equation}
Since $\phi_{-n}\ge 0$ for all $n$, this sequence must end. Namely,
for some $m$, $\phi_m =0$. Then $0 = \langle \phi_{-m}\vert \hat
H\vert \phi_{-m}\rangle = \Big\vert \tilde
a\vert\phi_{-m}\rangle\Big\vert^2\ge 0\>$ or $\tilde a\vert 
\phi_{-m}\rangle=0$. The ground state is then identified as
$\vert\Omega \rangle = \vert \phi_{-m}\rangle$ and
$H\vert\Omega\rangle=\epsilon_0 e_g \vert\Omega\rangle$.

The existence of a groundstate is only gaurenteed when $\lambda$ is
positive definite. In the semi-classical limit it can be shown that 
this holds for wide classes of bounding potentials. For the
polynomial interaction potential $V(a,a^\dagger) =  
\epsilon_1(a+a^\dagger)^l/l$, 
\begin{equation}
\lambda^{(sc)} \sim
e^{1/2-1/l}\left(\frac{\epsilon_1}{\epsilon_0}\right)^{1/l}\>.
\label{a4}
\end{equation}
For the exponential interaction potential $V(a,a^\dagger) =
\epsilon_1e^{\alpha^2(a+a^\dagger)^2}$, 
\begin{equation}
\lambda^{(sc)}
\sim\alpha\sqrt{\frac{e}{\log{(e\epsilon_0/\epsilon_1)}}}\>.
\label{a5}
\end{equation}
In both cases $\lambda^{(sc)}$ is positive definite. Notice also that
$\lambda^{(sc)}/e^{(sc)}\to 0$ as $e^{(sc)}\to \infty$, justifying the
approximations made in $\bf Sec V$.

\centerline{\bf Appendix B}

The form of eq.~$(\ref{e6})$ as well as the algebraic equation
eq.~$(\ref{n2})$ suggests that we look at a finite difference form of
the one dimensional Lie derivative. Given any function $\lambda(x)$,
we define the finite difference Lie operator ${\cal L}_\lambda$ by
\begin{equation}
{\cal L}_\lambda f(x) \equiv f(\lambda(x) + x) -f(x)\>,
\label{b1}
\end{equation}  
where $f(x)$ is any function of $x$. In the limit $\lambda(x)\to 0$
for all $x$, it is straightforward to see that eq.~$(\ref{b1})$ reduces
to the usual Lie derivative. Moreover, for any two functions $f(x)$
and $g(x)$ and constants $a,b$, 
\begin{equation}
{\cal L}_\lambda \left(af(x) + bg(x)\right) = a{\cal L}_\lambda f(x) +
b{\cal L}_\lambda g(x)\>.
\label{b2}
\end{equation}  
{}${\cal L}_\lambda$ is therefore a linear operator on the space of functions on
${\bf R}^1$. It is {\it not}, however, a derivation since it does not
satisfy the product rule,
\begin{equation}
{\cal L}_\lambda \left\{f(x)g(x)\right\} = 
	f(x){\cal L}_\lambda g(x)
	+
	g(x){\cal L}_\lambda f(x)
	+
	{\cal L}_\lambda f(x){\cal L}_\lambda g(x)\>.
\label{b3}
\end{equation}
Finally, for any two given functions $\lambda(x)$ and $\xi(x)$, the
commutator of two finite difference Lie operators 
\begin{equation}
[{\cal L}_\lambda, {\cal L}_\xi]f(x) =
f(x+\lambda(x)+\xi(x+\lambda(x)))-  f(x+\xi(x)+\lambda(x+\xi(x)))\>,
\label{b4}
\end{equation}
vanishes if and only if ${\cal L}_\lambda \xi(x) = {\cal L}_\xi
\lambda(x)\>$. 

Using ${\cal L}_\lambda$, the particular solution of eq.~$(\ref{n2})$
becomes the solution of the operator equation,
\begin{equation}
{\cal L}_\lambda N_p = 1\>.
\label{b5}
\end{equation}
Of more interest is the homogeneous solution to eq.~$(\ref{n2})$,
\begin{equation}
{\cal L}_\lambda N_h = 0\>,
\label{b6}
\end{equation}
which lies in the kernel of ${\cal L}_\lambda$, ${\rm ker}\> {\cal L}_\lambda$,
for a given $\lambda$. Notice that when $\lambda$ is a real
constant, ${\rm ker}\> {\cal L}_\lambda$ is the space of all periodic
functions with period $\lambda$. When $\lambda$ is a function of $x$,
${\rm ker}\> {\cal L}_\lambda$ will contain generalizations of periodic
functions to those whose frequencies are $x$ dependent. This agrees
quite well with the observation that in the semi-classical limit,
$1/\lambda^{(sc)}$ reduces to the WKB result. Of particular interest
is when $\lambda\in {\rm ker}\> {\cal L}_\lambda$:
\begin{equation}
{\cal L}_\lambda \lambda = 0\>.
\label{b7}
\end{equation}
From eq.~$(\ref{e10})$ we see that for this $\lambda$ the energy
levels are equally spaced and is determined solely by
$\lambda(e_g)$. The constant function, and thus the SHO, clearly
satisfies eq.~$(\ref{b7})$. Whether there exists other non-trivial
solutions to eq.~$(\ref{b7})$ for physically realizable non-linear
oscillators is still an open question.

\newpage


\begin{references}
\bibitem[29]{com2} The case where $V$ depends on time explicitly, as
	    through the coupling constant $\epsilon_1$, is much more
	    subtle. Neither eq.~$(\ref{e12})$ nor eq.~$(\ref{e15})$ are
	    valid since in general $[H(t),H(t')]\ne 0$ for $t\ne t'$.
\end{references}
\end{document}